\documentclass[prl,aps,twocolumn,groupedaddress,floats,showpacs,final,superscriptaddress]{revtex4-1}
\usepackage[detect-none]{siunitx}
\usepackage{ae}
\usepackage[T1]{fontenc}
\usepackage[ansinew]{inputenc}
\usepackage{amsmath}
\usepackage{graphicx}
\usepackage{color}
\usepackage[colorlinks]{hyperref}
\usepackage{setspace}
\begin{document}

\date{\today}
\title{Spin mediated photo-mechanical coupling of a nanoelectromechanical shuttle}

\author{A. V. Parafilo}
\affiliation{The Abdus Salam International Centre for Theoretical
Physics, Strada Costiera 11, I-34151 Trieste, Italy}

\author{S. I. Kulinich}
\affiliation{B. Verkin Institute for Low Temperature Physics and
Engineering of the National Academy of Sciences of Ukraine, Prospekt Nauky 47, Kharkov 61103, Ukraine}

\author{L. Y. Gorelik}
\affiliation{Department of %Applied
Physics, Chalmers University of
Technology, SE-412 96 G{\" o}teborg, Sweden}

\author{M. N. Kiselev}
\affiliation{The Abdus Salam International Centre for Theoretical
Physics, Strada Costiera 11, I-34151 Trieste, Italy}
\author{R. I. Shekhter}
\affiliation{Department of Physics, University of Gothenburg, SE-412
96 G{\" o}teborg, Sweden}
\author{M. Jonson}
\affiliation{Department of Physics, University of Gothenburg, SE-412
96 G{\" o}teborg, Sweden} \affiliation{SUPA, Institute of Photonics
and Quantum Sciences, Heriot-Watt University, Edinburgh EH14 4AS,
Scotland, UK}

\date{\today}
\pacs{85.85.+j, 85.75.-d, 73.23.Hk}

\begin{abstract}
We show that nano-mechanical vibrations in a magnetic shuttle device can be strongly affected by external microwave irradiation through photo-assisted electronic spin-flip transitions. Mechanical consequences of these spin-flips are due to a spin-dependent magnetic force, which may lead to a nano-mechanical instability in the device. We derive a criterion for the instability to occur and analyze different regimes of nano-mechanical oscillations. Possible experimental realizations of the spin-mediated photo-mechanical instability and detection of the device back action are discussed.
\end{abstract}
\maketitle

The electric charge of electrons injected by tunneling into
a nano-device provides a means for coupling mechanical deformations
of the device to electronic degrees of freedom. Different scenarios of
nano-electro-mechanical (NEM) action,
which provide a number of new
nano-device functionalities,
are based on a coupling via
either the injected electrical current or via charge accumulated
in the device (see, e.g., the reviews \cite{rev1, rev2, rev3}).
The electronic spin, which is usually almost decoupled from the charge
degrees of freedom in 
bulk nonmagnetic metals, might also contribute
to the mechanics of nanometer-sized devices. 
Reasons for this can be the amplified spin-orbit interaction in low dimensional non-magnetic
conductors \cite{markus, ahar, rastelli} or the
exchange interaction induced in magnetic NEM structures
\cite{pulkin, atal}.

Microwave electromagnetic fields are in general not expected to affect
the mechanical operation of a nanodevice very much. This is because of the
considerable mismatch between the vibration frequency of a typical nanomechanical
resonator ($\sim$100~MHz) and the frequency of an electromagnetic field
in the microwave/far infrared region (0.1-1~THz).
In this Letter, however, we show theoretically that the electronic spin
accumulated in a mechanical resonator can
mediate a strong coupling between a high-frequency electromagnetic field and
low-frequency mechanical vibrations in
a magnetic NEM system such as the one sketched in Fig.~1(a). The device shown
there comprises a single-wall carbon nanotube (CNT) resonator suspended
between ferromagnetic source- and drain electrodes and a magnetic gate.
The magnetization of the two electrodes are assumed to be antiparallel,
while the magnetization of the gate is taken to be antiparallel to that of the source.

%%%%%%%%%%%%%%%%%%%%%%%%%%%%%%%%%%%%%%%%%%%%%%%%%%%%%%%%
\begin{figure}
\centering
\includegraphics[width=0.95\columnwidth]{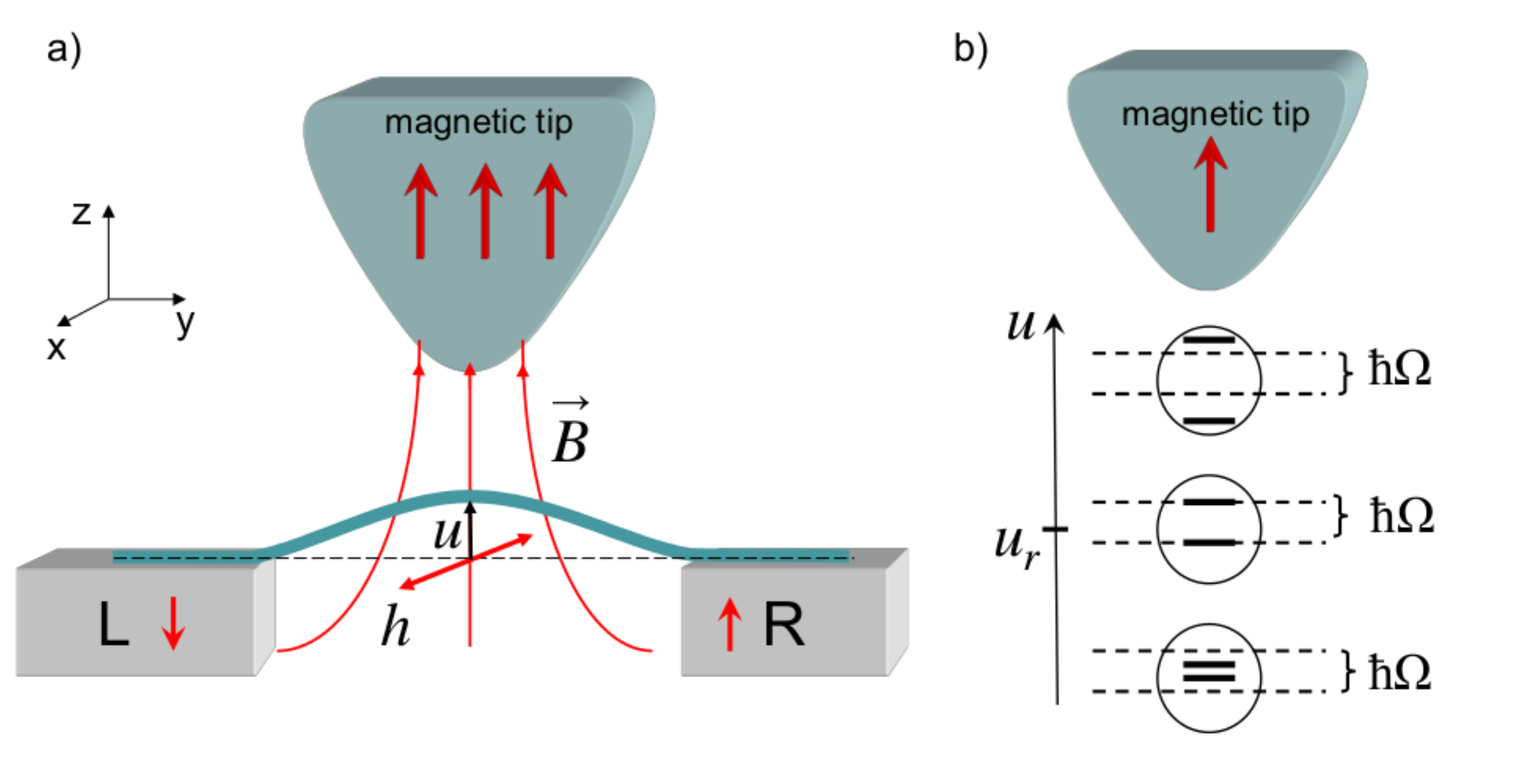}
\caption{ a) Sketch of a device with a suspended CNT resonator in
tunneling contact with two ferromagnetic
electrodes (L,R) with opposite magnetization directions.
Flexural CNT vibrations are actuated by a magnetic force caused by
the Zeeman interaction between the spin of CNT electrons and an
inhomogeneous magnetic field $B_z(z)$ from a magnetic gate (tip). b)
A microwave field $h\cos(\Omega t)$ induces electron spin-flips in
the CNT, which are resonant at $u$$=$$u_r$ where %if
the photon energy $\hbar\Omega$
equals the deflection-dependent Zeeman energy splitting
$g\mu_BB_z(u)$. Since the magnetic force is spin-dependent, the spin
in this way mediates the photo-mechanical coupling discussed in the
text. } \label{Fig1}
\end{figure}
%%%%%%%%%%%%%%%%%%%%%%%%%%%%%%%%%%%%%%%%%%%%%%%%%%%%%%%%%%

The inhomogeneous magnetic field created by the tip-shaped gate gives
rise to a Zeeman splitting of the CNT electronic energy levels that varies
with the deflection of the CNT [see Fig.~1(a,b)] and hence corresponds to a
magnetic force that depends on the net spin accumulated
in the CNT. This is the mechanism for how spin couples to the CNT
vibrations.
Turning to the mechanism for how spin couples to the
microwave field, we assume that the CNT is connected to the electrodes by
high-resistance tunnel barriers and that it is short enough ($\lesssim $$ 1$$~$$\mu$m)
for the spatial quantization of energy levels to be resolved and that we are
in the Coulomb blockade regime of transport at low temperatures ($\sim $$ 1$~K).
This means that we can treat the CNT as a quantum dot (QD) with a
single electron energy level, doubly degenerate due to the spin
degree of freedom at zero magnetic field but Zeeman split at finite fields.
The coupling of the microwave
field to the electronic spin is through
photo-assisted spin-flip scattering, which
becomes resonant if the microwave photon energy equals the Zeeman splitting
[see Fig.~1(b)]. Each spin-flip changes the amount of spin accumulated in the CNT/QD in
the sense that the value of the spin projection on the axis defined by the co-linear
magnetizations of electrodes and gate is changed by one unit, effectively adding an
``extra spin" to the CNT resonator. Such photo-induced ``spin pumping"
 also changes the spin-dependent magnetic force on the CNT. This, in qualitative terms,
 is how the electron spin can mediate an interaction between nanomechanical
 vibrations and a microwave 
 field. To see what consequences such a coupling may have
 we will ask the question whether the magnetic force does positive (amplification) or
 negative (damping) work on the CNT resonator during one oscillation period.
 The answer turns out to depend on the ``sign" of the extra spin, which
 in its turn depends on which of the two oppositely magnetized electrodes is the source
 of spin-polarized electrons injected into the CNT/QD. But this can be changed by
 changing the polarity of the voltage bias applied between the two ferromagnetic
 electrodes. Hence we can be certain that for either positive or negative polarity
 energy will be pumped into the vibrations of the CNT resonator giving
 rise to a mechanical instability and to pronounced vibrations if the pumping can
 overcome the dissipation in the mechanical subsystem.

In order to rigorously demonstrate the phenomenon that we so far have only described
qualitatively for our model system we assume that the dynamics of the CNT/QD resonator
is completely characterized by the amplitude of its fundamental bending mode $u(t)$,
whose time evolution is governed by %the classical
Newton's equation for a harmonic oscillator of eigenfrequency $\omega_0$,
%\vspace*{-5mm}
%%%%%%%%%%%%%%%%%%%%%%%%%%%%%%%%%%%%%%%%%%%%%%%%%%%%%%%%%%%%%%%%%%%%%%%%%%%%%%%%%%%%%%
\begin{eqnarray}\label{EOM}
%\ddot{u}+\gamma \dot{u}+\omega_0^2 u = \frac{F}{ m} {\rm Tr}
m\left[ \ddot{u}+\gamma \dot{u}+\omega_0^2 u\right] = F\,{\rm Tr}
%\ddot{u}+\gamma \dot{u}+\omega_0^2 u = -(F/m) {\rm Tr}
\left\{\hat{\rho}\left(\hat{n}_{\downarrow}-\hat{n}_{\uparrow}\right)\right\}
+F_{\rm elast}^{\rm eq}.
%\left\{\hat{\rho}\left(\hat{n}_{\downarrow}-\hat{n}_{\uparrow}\right)\right\}.
\end{eqnarray}
%%%%%%%%%%%%%%%%%%%%%%%%%%%%%%%%%%%%%%%%%%%%%%%%%%%%%%%%%%%%%%%%%%%%%%%%%%%%%%%%%%%%%%%%
Here $\gamma$$=$$\omega_0$$/$$Q_0$ is a phenomenological damping rate,
$Q_0$ is the mechanical quality factor and $m$ is the effective mass of the resonator.
The first force term on the r.h.s, % of Eq.~(\ref{EOM}),
where $F$$=$$($$g$$\mu_B$$/$$2$$)$$\partial B_z(u)$$/$$\partial u
|_{u=0}$
%{\color{blue} here I add footnote}
%%\cite{note}
and
$\hat{n}_{\sigma}$$=$$d^{\dag}_{\sigma}d_{\sigma}$ is the density
operator
for QD electrons with spin $\sigma$$=$$\uparrow, \downarrow$, is the %spin-mechanical
magnetic force induced by the
interaction between the QD's spin and the gate-induced inhomogeneous magnetic field $B_z(u)$,
which is assumed to have a linear dependence on $u$ in the region of interest. %$g$ is the gyromagnetic
%ratio, $\mu_B$ is the Bohr magneton,
%$m$ is the CNT's mass, \mats{and}
%$\hat{n}_{\sigma}=d^{\dag}_{\sigma}d_{\sigma}$ is the density operator
%for QD electrons with spin $\sigma=\uparrow, \downarrow$ %localized in the QD
%\mats{Can we delete the following parenthesis? ($d_{\sigma}, d_{\sigma}^{\dag}$ are annihilation and creation
%operators, respectively)}.
The second force term is the elastic restoring force that
compensates the magnetic force in the absence of microwave
irradiation introduced so that the %midpoint
deflection coordinate
$u$ (along the $z$-axis, see Fig.~1) is measured from the midpoint
of the (bent) resonator in this static case.
The trace %on the r.h.s. of
in Eq.~(\ref{EOM}) is over the electronic degrees of freedom. Consequently,
$\hat{\rho}$ is the electron density matrix operator, which obeys %a
the quantum Liouville-von Neumann equation %that reads
($\hbar$$=$$1$)
%%%%%%%%%%%%%%%%%%%%%%%%%%%%%%%%%%%%%%%%%%%%%%%%%%%%%%%%%%%%%%%%%%%%%%%%%%%%%%%%%%%%%%
\begin{eqnarray}\label{LvN}
i\frac{d \hat{\rho}}{d t}=[\hat{H}_d+\hat{H}_l+\hat{H}_t,
\hat{{}\rho}].
\end{eqnarray}
%%%%%%%%%%%%%%%%%%%%%%%%%%%%%%%%%%%%%%%%%%%%%%%%%%%%%%%%%%%%%%%%%%%%%%%%%%%%%%%%%%%%%%
 Here $\hat{H}_d$ is the QD Hamiltonian,
%%%%%%%%%%%%%%%%%%%%%%%%%%%%%%%%%%%%%%%%%%%%%%%%%%%%%%%%%%%%%%%%%%%%%%%%%%%%%%%%%%%%%%%
\begin{eqnarray}\label{Model}
&&\hat{H}_d=\varepsilon_0
(\hat{n}_{\uparrow}+\hat{n}_{\downarrow})+\frac{g \mu_B
B_z(u)}{2}(\hat{n}_{\uparrow}-\hat{n}_{\downarrow})
\nonumber \\
&&+\frac{g \mu_B h}{2}\cos(\Omega
t)(d^{\dag}_{\uparrow}d_{\downarrow}+d^{\dag}_{\downarrow}d_{\uparrow})-
Ud_{\uparrow}^{\dag}d^{\dag}_{\downarrow}d_{\uparrow}d_{\downarrow},
\end{eqnarray}
%%%%%%%%%%%%%%%%%%%%%%%%%%%%%%%%%%%%%%%%%%%%%%%%%%%%%%%%%%%%%%%%%%%%%%%%%%%%%%%%%%%%%%%%%
where $\varepsilon_0$ is the energy of the QD level relative to the
Fermi energy in the leads, {$g \mu_B B_z(u)$ is the CNT's
deflection-dependent Zeeman energy splitting, $h$ is the amplitude of the
magnetic component of the external microwave field, which oscillates with frequency $\Omega$
along the $x$-axis [see Fig. 1(a)]},
and $U$ is the Coulomb (charging) energy cost for double occupancy of the QD.
%electron Coulomb (charging) energy.
Assuming that the variation of
the %longitudinal
magnetic field $B_z$ along the mechanical
motion is small compared to its value at the equilibrium position of the dot, $\Delta$$=$$g\mu_B$$
B_z($$u$$=$$0)$, we retain only the linear term in its Taylor expansion,
$g \mu_B B_z(u)$$=$$\Delta$$+$$2 F%\cdot
u$.

The second term in the commutator in Eq.~(\ref{LvN})
%, \mats{Delete this equation?}
%%%%%%%%%%%%%%%%%%%%%%%%%%%%%%%%%%%%%%%%%%%%%%%%%%%%%%%%%%%%%%%%%%%%%%%%%%%%%%%%%%%%%%%%%
%\begin{eqnarray}\label{lead}
%\hat{H}_l=\sum_{k}\varepsilon_k\left(c_{L k\uparrow }^{\dag}c_{L
%k\uparrow }+c_{R k\downarrow }^{\dag}c_{R k\downarrow }\right),
%\end{eqnarray}
%%%%%%%%%%%%%%%%%%%%%%%%%%%%%%%%%%%%%%%%%%%%%%%%%%%%%%%%%%%%%%%%%%%%%%%%%%%%%%%%%%%%%%%%%
describes non-interacting %spin-polarized
electrons %with energy $\varepsilon_k$
in the %left (L) and right (R)
electrodes (for now assumed to be 100\% spin polarized ``half metals" \cite{halfmetals})
%{\color{blue}(initially we assume $100\%$ polarization in
%the leads []? I can put here the cite on exp. work, the same that
%you used in your recent PRL about magnetic shuttling)},
while the
%, where $c_{jk\sigma}$ are $c_{jk\sigma}^{\dag}$ are annihilation and creation electron operators, $j=L,R$. The
third term, %in the commutator,
%%%%%%%%%%%%%%%%%%%%%%%%%%%%%%%%%%%%%%%%%%%%%%%%%%%%%%%%%%%%%%%%%%%%%%%%%%%%%%%%%%%%%%%%%%
\begin{eqnarray}\label{tunneling}
%\hat{H}_t=\sum_{k\sigma} [\mathcal{T} (e^{ieVt/2} c^{\dag}_{L k
%\uparrow } d_{\uparrow}+e^{-ieVt/2} c^{\dag}_{R k \downarrow }
%d_{\downarrow})+ h.c.]\,,
\hat{H}_t=\sum_{k} \mathcal{T} [e^{ieVt/2} c^{\dag}_{L k
\uparrow } d_{\uparrow}+e^{-ieVt/2} c^{\dag}_{R k \downarrow }
d_{\downarrow}]+ h.c.,
\end{eqnarray}
%%%%%%%%%%%%%%%%%%%%%%%%%%%%%%%%%%%%%%%%%%%%%%%%%%%%%%%%%%%%%%%%%%%%%%%%%%%%%%%%%%%%%%%%%%%
is a standard tunnel Hamiltonian. Here %representing
%with
electron tunneling
is characterized by the
%overlapping
overlap integral  $\mathcal{ T}$, %and with
a bias voltage $V$ is
applied between the electrodes,
and $c_{jk\sigma}^{\dag}$ are electron creation operators acting in the left ($j$$=$$L$) and right ($j$$=$$R$) electrodes.
%Without loss of generality
For simplicity %put
we take the %tunneling
amplitudes for tunneling to the left and right %equal to each other.
to be the same \cite{footnote2}. %We denote the left electrode as a source.
The
%in the source (left) electrode and in the
%magnetic tip are assumed to be parallel to each other.
magnetization in the source (left if $V$$>$$0$) electrode and
in the magnetic tip are assumed to be antiparallel
 to each other \cite{directions}.
%in Eqs.~(\ref{lead}-\ref{tunneling}). \mats{Do we need the next sentence?}
To investigate the %case of antiparallel magnetization one should
%antiparallel
parallel case one has to reverse the polarity of the bias voltage, %replace
$V$$\rightarrow$$ -$$V$, in Eq.~(\ref{tunneling}), so that the right electrode becomes the source.
%the polarity of the bias voltage $eV\rightarrow-eV$.

By treating the tunnel Hamiltonian perturbatively and using the reduced
density matrix approach \cite{novotny,fedorets} one can
factorize the density operator,
%%%%%%%%%%%%%%%%%%%%%%%%%%%%%%%%%%%%%%%%%%%%%%%%%%%%%%%%%%%%%%%%%%%%%%%%%%%%%%%%%%%%%%%%%%%%%%
%\begin{equation}\label{factor}
$\hat{\rho}(t)$$=$$\hat{\varrho}_d(t)\otimes\hat{\rho}_{l}$,
%\end{equation}
%%%%%%%%%%%%%%%%%%%%%%%%%%%%%%%%%%%%%%%%%%%%%%%%%%%%%%%%%%%%%%%%%%%%%%%%%%%%%%%%%%%%%%%%%%%%%%%
into a product of an equilibrium density operator for the leads at temperature $T$,
$\hat{\rho}_l$$=$$\exp(-\hat{H}_l/k_B T)$,
%($k_B$ is the Boltzmann constant, $T$ is the temperature),
and a reduced density operator for the QD states,
%%%%%%%%%%%%%%%%%%%%%%%%%%%%%%%%%%%%%%%%%%%%%%%%%%%%%%%%%%%%%%%%%%%%%%%%%%%%%%%%%%%%%%%%%%%%%%%%%
\begin{eqnarray}\label{rhod}
\hat{\varrho}_d=c_0|0\rangle\langle0|+c_2 |\uparrow\downarrow\rangle\langle \uparrow\downarrow|
+|\sigma\rangle\varrho_{\sigma\sigma'}\langle\sigma'|.
\end{eqnarray}
%%%%%%%%%%%%%%%%%%%%%%%%%%%%%%%%%%%%%%%%%%%%%%%%%%%%%%%%%%%%%%%%%%%%%%%%%%%%%%%%%%%%%%%%%%%%%%%%%
%where
In Eq. (\ref{rhod}) $|\sigma\rangle$$=$$d^{\dag}_{\sigma}|0\rangle$,
$|\uparrow\downarrow\rangle$$=$$d^{\dag}_{\uparrow}d^{\dag}_{\downarrow}|0\rangle$ are
singly and doubly occupied electron states,
$\varrho_{\sigma\sigma'}$$=$$\langle\sigma|\hat{\varrho}_d|\sigma'\rangle$
and $c_0$, $c_2$ are normalization constants. We consider a NEM system
in the strong Coulomb blockade regime, $U$$\gg$$|eV/2\pm\varepsilon_0|$, for which
%the electron
double electron occupancy of the dot is forbidden, so that $c_2$$=$$0$. By averaging
over the electronic states in the leads we get an equation for the
matrix
%%%%%%%%%%%%%%%%%%%%%%%%%%%%%%%%%%%%%%%%%%%%%%%%%%%%%%%%%%%%%%%%%%%%%%%%%%%%%%%%%%%%%%%%%%%%%%%%%%%%%%%%%%%%%%%%%%%%%%%%%%%%%%%%%%
\begin{eqnarray}\label{dmo}
\hat{\varrho}=\left(\begin{array}{cc} \varrho_{\uparrow\uparrow}& \varrho_{\uparrow\downarrow} \\

                          \varrho_{\downarrow\uparrow}&
                         \varrho_{\downarrow\downarrow}
                         \end{array}\right)\,
\end{eqnarray}
%%%%%%%%%%%%%%%%%%%%%%%%%%%%%%%%%%%%%%%%%%%%%%%%%%%%%%%%%%%%%%%%%%%%%%%%%%%%%%%%%%%%%%%%%%%%%%%%%%%%%%%%%%%%%%%%%%%%%%%%%%%%%%%%%
which characterizes the state of the singly occupied QD level. This
equation reads
%%%%%%%%%%%%%%%%%%%%%%%%%%%%%%%%%%%%%%%%%%%%%%%%%%%%%%%%%%%%%%%%%%%%%%%%%%%%%%%%%%%%%%%%%%%%%%%%%%%%%%%%%%%%%%%%%%%%%%%%%%%%%%
\begin{eqnarray}\label{dmsys}
\dot{\hat{\varrho}}=-\frac{i}{2}[(\Delta+2F%\cdot
u)\sigma_z+\nu_h\cos(\Omega
t)\sigma_x,\hat{\varrho}]-\frac{\Gamma}{2}\mathfrak{L}[\hat{\varrho}],
\end{eqnarray}
%%%%%%%%%%%%%%%%%%%%%%%%%%%%%%%%%%%%%%%%%%%%%%%%%%%%%%%%%%%%%%%%%%%%%%%%%%%%%%%%%%%%%%%%%%%%%%%%%%%%%%%%%%%%%%%%%%%%%%%%%%%%%%%%
where
%%%%%%%%%%%%%%%%%%%%%%%%%%%%%%%%%%%%%%%%%%%%%%%%%%%%%%%%%%%%%%%%%%%%%%%%%%%%%%%%%%%%%%%%%%%%%%%%%%%%%%%%%%%%%%%%%%%%%%%%%%%%%%%%%%%%%%%
\begin{eqnarray}\label{Lind}
&&\mathfrak{L}[\hat{\varrho}]=\frac{3}{2}\hat{\varrho}+\frac{1}{2}\sigma_z\hat{\varrho}\sigma_z+
2\sigma_{\kappa}\hat{\varrho}\sigma_{-\kappa}-(\hat{1}+\kappa\sigma_z).
\end{eqnarray}
%%%%%%%%%%%%%%%%%%%%%%%%%%%%%%%%%%%%%%%%%%%%%%%%%%%%%%%%%%%%%%%%%%%%%%%%%%%%%%%%%%%%%%%%%%%%%%%%%%%%%%%%%%%%%%%%%%%%%%%%%%%%%%%%%%%%
Here $\sigma_{x,y,z}$ %$\vec{\sigma}$ stands for
are the Pauli matrices,
$\sigma_{\pm}$$=$$(\sigma_x$$\pm$$ i\sigma_y)/2$, $\nu_h$$=$$g\mu_B
h$, %and
$\Gamma$$=$$2 \pi \nu |\mathcal{T}|^2$ is the energy
level width, and $\nu$ is the density of states %(DoS)
in the leads. In Eq.~(\ref{Lind}) $\kappa$$=$${\rm sign}(eV)$ determines
the relative
%spin-polarization orientations
magnetization in the source and the tip: $\kappa$$=$$+1$
stands for
%parallel,
antiparallel, while $\kappa$$=$$-1$ stands for
%antiparallel
parallel alignment. The first term in
Eq.~(\ref{dmsys}) describes the dynamical part of the time evolution
of the electronic subsystem, while the second term is a stochastic
part associated with tunneling processes from/to the leads. We
assume the condition of a large bias voltage,
$|eV|$$\gg$$($$k_B$$T$$,$$\Delta$$,$$\Omega$$,$$\Gamma)$.

Having in mind microwave-field intensities that can be achieved experimentally
in the THz frequency range,
%achievable intensities  fields of the terahertz frequencies
we will consider a situation where $\nu_h$$\ll$$ \Omega $$\sim$$ \Delta$.
Under this condition
one can apply the rotating wave approximation, which makes it possible to %allows to
remove the
explicit time dependence in Eq.~(\ref{dmsys}) by a unitary
transformation. As a result,
we may replace $\nu_h\cos(\Omega t)$ by $\nu_h/2$ and
renormalise $(\Delta$$+$$2F%\cdot
u)$$\rightarrow$$(\Delta$$-$$\Omega$$+$$2Fu)$  in Eq.~(\ref{dmsys}). We then evaluate Eq.~(\ref{dmsys})
under the assumption that the motion of the  QD is adiabatic %motion
($\omega_0$$\ll$$\Gamma$), so that the characteristic times associated with mechanical motion
and electron tunneling are well separated. It follows that the equation of
motion %(EoM)
(\ref{EOM}) for a small-amplitude microwave field,
$\nu_h$$\lesssim$$ \Gamma$, takes the form
%\vspace*{-3mm}
%%%%%%%%%%%%%%%%%%%%%%%%%%%%%%%%%%%%%%%%%%%%%%%%%%%%%%%%%%%%%%%%%%%%%%%%%%%
\begin{eqnarray}\label{newteq1}
%&& \ddot{\tilde{u}}+\gamma \dot{\tilde{u}} + \omega_0^2
%\tilde{u}=w_1^2f_1(\tilde{u})-\frac{w_2^2}{\Gamma}f_2(\tilde{u})\dot{\tilde{u}},
&& \ddot{u}+\gamma \dot{u} + \omega_0^2 u =
w_1^2f_1(u)+\frac{w_2^2}{\Gamma}f_2(u)\dot{u}\,,
\end{eqnarray}
%%%%%%%%%%%%%%%%%%%%%%%%%%%%%%%%%%%%%%%%%%%%%%%%%%%%%%%%%%%%%%%%%%%%%%%%%%%%%
where %$\tilde{u}$$=$$u$$-$$u_s$
(as before) $u$ is the CNT's deflection relative to its
equilibrium position in the absence of a microwave field, %$u_s$$=$$-$$\kappa F$$/m\omega_0^2$,
%and
%when microwave field is absent,
%\vspace*{-3mm}
%%%%%%%%%%%%%%%%%%%%%%%%%%%%%%%%%%%%%%%%%%%%%%%%%%%%%%%%%%%%%%%%%%%%%%%%%%%%%%%%%%%%%%%%%
\begin{eqnarray}\label{freq}
w_1^2=\frac{3 F\nu_h^2}{4m\Gamma^2} \quad,\quad w_2^2=\frac{4F^2\nu_h^2}{m \Gamma^3} \,,
%w_1=\sqrt{3F/4m\Gamma^2} \,\nu_h\,;\,\,
%w_2=\sqrt{4 F^2 / m \Gamma^3}\, \nu_h
\end{eqnarray}
%%%%%%%%%%%%%%%%%%%%%%%%%%%%%%%%%%%%%%%%%%%%%%%%%%%%%%%%%%%%%%%%%%%%%%%%%%%%%%%%%%%%%%%%%%
and %\vspace*{-2mm}
%%%%%%%%%%%%%%%%%%%%%%%%%%%%%%%%%%%%%%%%%%%%%%%%%%%%%%%%%%%%%%%%%%%%%%%%%%%%%%%%%%%%%%%
%\begin{eqnarray}\label{not}
%&&
%f_1(\tilde{u})=\frac{\kappa}{1+4J^2(\tilde{u})},\;
%f_2(\tilde{u})=\kappa
%J(\tilde{u})\frac{4J^2(\tilde{u})+13}{(1+4J^2(\tilde{u}))^3}
%\end{eqnarray}
\begin{equation}\label{not}
f_1(u)=\frac{\kappa}{1+4J^2(u)};\;
f_2(u)=\kappa
J(u)\frac{4J^2(u)+13}{[1+4J^2(u)]^3}\,,
\end{equation}
%%%%%%%%%%%%%%%%%%%%%%%%%%%%%%%%%%%%%%%%%%%%%%%%%%%%%%%%%%%%%%%%%%%%%%%%%%%%%%%%%%%%%%%
%Here we use the notation
where
%$J(\tilde{u}) = (\tilde{\Delta} - \Omega - 2 F%\cdot
%\tilde{u})/\Gamma$, where $\tilde{\Delta} = \Delta - 2F% \cdot $$
%u_s$
$J(u)$$ = $$(\Delta $$-$$ \Omega $$+ $$2 F%\cdot
u)/\Gamma$.
%, where $\tilde{\Delta} = \Delta - 2F% \cdot $$
%u_s$
%is the renormalized Zeeman energy splitting.
The function $f_1(u)$ describes a nonlinear force, %that
%acts on the CNT %due to
%through the SMC \mats{(Rewrite to avoid "SMC"?)} and
%renormalizes the vibration frequency of the resonator,
while the function
$f_2(u)$ is a nonlinear friction term. %, which %is determined by
Both functions depend on the spin accumulated on the QD due to photo-induced
electronic spin-flip transitions; hence their values vary rapidly as the
deflection $u$ of the CNT/QD resonator is close to satisfying the resonance
condition $J$$($$u$$=$$u_r$$)$$=$$0$, where the rate of spin-flip transitions is maximal.

Whether or not microwave induced spin-flip transitions will
contribute  a ``negative friction" $(f_2$$>$$0)$, possibly leading
to a nanomechanical instability, depends, as will be shown by the
stability analysis that follows, on whether resonant spin-flip
transitions occur for a deflection towards ($u_r$$>$$0)$ or away
from ($u_r$$<$$0$) the magnetic gate.

%The relative position of the resonant point $u=u_r$ with respect to equilibrium position of the dot $u=0$ determines the possibility to activate a ``negative" friction $f_2 $$< 0$ caused by microwave radiation which in its turn may cause a nanomechanical instability. This can be found by analysis of the condition for shuttle instability.}

%%\vspace*{-3mm}
%%At first we
%We first %investigate
%find the criterion for a shuttling instability to occur, a criterion that is
%determined by the conditions under which the %eigenfrequency
%\mats{frequency $\omega$} of the \mats{forced} CNT %oscillator
%\mats{oscillations}
%develops a negative imaginary part.
%%of appearance of a negative imaginary
%%correction to the eigenfrequency of the oscillator.

Linearizing %the EoM
Eq.~(\ref{newteq1}) assuming $\vert u\vert$$\ll$$ {\rm Max}\left\{ \vert u_r\vert,\Gamma/F \right\}$  %and %
corresponds to expanding $f_1(u)$ to first and $f_2(u)$ to zeroth order in $u$.
Neglecting the former terms, %dropping %off  the
%terms %linear in the displacement,
which lead to a %are % terms responsible for a
small renormalization of the static shift and vibration frequency of the resonator,
%shift $\delta \omega_0\ll\omega_0$ in the vibration frequency, %mechanical frequency shift
%($\delta \omega_0\ll\omega_0$). Keeping
%and keeping only the \mats{0:th term in the expansion of $f_2(u)$} %terms proportional to $\dot{u}$,
we obtain %the equation
an expression for the imaginary part of
the eigenfrequency, % that reads
\begin{equation}
%{\rm Im}[\omega]=(\gamma-\gamma_h)/[2+(\gamma-\gamma_h)^2/2],
%{\rm Im}[\omega]\propto(\gamma-\gamma_h)/2\,,
{\rm Im}[\omega] = (\gamma-\gamma_h)/2\,,
\end{equation}
where $\gamma_h $$\propto$$ f_2(0)$ and
%\vspace*{-3mm}
%%%%%%%%%%%%%%%%%%%%%%%%%%%%%%%%%%%%%%%%%%%%%%%%%%%%%%%%%%%%%%%
%\begin{eqnarray}\label{gamma}
%{\rm Im}[\omega]&=&(\gamma-\gamma_h)/[2+(\gamma-\gamma_h)^2/2], \\
%\gamma_h&=&\kappa(\Omega-\tilde{\Delta})w^2_2\Gamma^2\frac{4(\tilde{\Delta}-\Omega)^2
%+13\Gamma^2}{[4(\tilde{\Delta}-\Omega)^2+\Gamma^2]^3} \nonumber
%\end{eqnarray}
\begin{equation}\label{gamma}
%\gamma_h=\kappa(\Omega-\tilde{\Delta})w^2_2\Gamma^2\frac{4(\tilde{\Delta}-\Omega)^2
%+13\Gamma^2}{[4(\tilde{\Delta}-\Omega)^2+\Gamma^2]^3}
\gamma_h=\kappa(\Delta-\Omega)w^2_2\Gamma^2\frac{4(\Delta-\Omega)^2
+13\Gamma^2}{[4(\Delta-\Omega)^2+\Gamma^2]^3}
\end{equation}
%%%%%%%%%%%%%%%%%%%%%%%%%%%%%%%%%%%%%%%%%%%%%%%%%%%%%%%%
is the microwave%magnetic-field
-induced ``friction" coefficient. We note that the
amplitude of the mechanical oscillations increases with time if
%develops when
$\gamma_h$$>$$\gamma$, corresponding to a mechanical ``shuttling" instability. To determine the %shuttling
instability criterion
we neglect %all effects of
the intrinsic friction
($\gamma$$=$$0$) of the mechanical subsystem. Then, %the case
 %{\color{blue}
 for antiparallel magnetization between tip and source ($\kappa$$=$$+$$1$),
%{\color{blue} (AP: note, due to changing of interaction term
%(magnetic field with magnetization, its another sign in Eq.(3)) our
%criterion for shuttling instability
%$\Omega>\Delta$ is actual for  the case of parallel magnetizations in source and tip, and its corresponds to $\kappa=-1$)}
shuttling %then
occurs when the %frequency of the microwave field exceeds the
microwave frequency is lower than %exceeds
the %value of the
Zeeman energy splitting,
%$\Omega>\tilde{\Delta}$
$\Omega$$<$$\Delta $. Therefore, whether an instability occurs or not %the oscillation instability
depends on %the location of the CNT's deflection %$u_r=\kappa(\tilde{\Delta}-\Omega)/2F$
whether the deflection $u_r=(\Omega-\Delta)/2F$ that corresponds to
resonant spin-flip transitions is in the direction of the magnetic
gate ($u_r$$>$$0$) or away from it ($u_r$$<$$0$). The latter case gives rise
to an instability, while the former leads to additional damping and
stability.
The corresponding criterion for parallel magnetizations
($\kappa$$=$$-$$1$) can be found from Eq.~(\ref{gamma}) in a similar
manner.

The instability threshold for %the
CNT oscillations is conditioned by
the equality of the microwave field-dependent contribution to the
quality factor, $Q_h$$=$$\omega_0$$/$$\gamma_h$, and the intrinsic mechanical
quality factor $Q_0$. This criterion determines the critical value
for the amplitude of the microwave field. Estimating $Q_h$ %$Q_h^{-1}$
by using the maximum value of the increment
($|\Delta$$-$$\Omega|$$\sim$$\Gamma$) given by Eq. (\ref{gamma}) we find
that
%\vspace*{-3mm}
%%%%%%%%%%%%%%%%%%%%%%%%%%%%%%%%%%%%%%%%%%%%%%%%%%%%%%%%%%%%%%%%%%%%%%%%
\begin{eqnarray}\label{qual}
Q_h^{-1}\sim \frac{1}{\omega_0} \frac{w_2^2}{\Gamma}\sim
u_0^2\frac{F^2}{\Gamma^2}\frac{\nu^2_{h}}{\Gamma^2},
\end{eqnarray}
%%%%%%%%%%%%%%%%%%%%%%%%%%%%%%%%%%%%%%%%%%%%%%%%%%%%%%%%%%%%%%%%%%%%%%%%
where $u_0$$=$$($$2$$m$$\omega_0$$)$$^{-1/2}$ $\simeq2$~pm (see, e.g.,
\cite{zantnano}) is the amplitude of the CNT's zero-point
oscillation. 
Using the realistic values 1~T for the magnetic field $B_z(0)$ and $5\cdot10^6$~T/m for the
field gradient $\partial B_z(0)/\partial z $ \cite{exp1,exp2}, while taking
the CNT quality factor $Q_0$ and the tunnel coupling
$\Gamma$ to be $10^6$ \cite{qualfact} and %{\color{blue}$0.7\cdot10^{-2}$~K}
$1\cdot10^9\,$s$^{-1}$ 
%[rad/s?; $0.9\cdot10^9$ (which corresponds to 7 mK)? $0.6\cdot10^9$ (used in Fig. 2)? ]} 
\cite{footnote2}, respectively,
one then finds %finally get
%\mats{for a CNT resonator of length $L\sim 100$~nm}
a lower bound of
$h$$ \sim$$1$~mT
for the magnetic
amplitude of the microwave field.
%This value also justifies the use of
%%gives us a justification for use of the
%perturbation theory ($\nu_h/\Gamma$$\sim
%$$10^{-1}$$ \ll $$1$) for solving %applied for the solution of the
%Eq.~(\ref{newteq1}).

 A partial rather than complete spin polarization of the leads would not 
 qualitatively change the threshold for shuttling since it only produces quantitative changes in the %DoS
leads' densities of state and modifies
$\Gamma^{\uparrow,\downarrow}_{L(R)}$. As a result, direct
%processes of electron transfer
electron transfer processes between the left and right electrodes
without spin-flips become possible, hence reducing the spin-flip rate. The
magnitude of the field induced friction is proportional to the
degree of spin %magnetic
polarization
$\eta$$=$$|\Gamma^{\uparrow}$$-$$\Gamma^{\downarrow}|/(\Gamma^{\uparrow}$$+$$\Gamma^{\downarrow})$,
where %we denote
$\Gamma^{\uparrow(\downarrow)}$$=$$\Gamma^{\uparrow(\downarrow)}_L$$=$$\Gamma^{\downarrow(\uparrow)}_R$.
Therefore, for partially polarized leads ($\eta$$<$$1$) the rate at which the
oscillation amplitude grows if the shuttle instability criterion is met
%increment of shuttle instability
decreases with $\eta$ as $\gamma_h$$\propto$$\eta(1$$+$$\eta)$ and vanishes
in the limit $\eta$$\rightarrow$$0$.

If there is
%If the conditions for
a shuttle instability %are met
the amplitude of the resonator vibrations increases with
time and we have to consider the full non-linear version of the equation of motion (\ref{newteq1}).
In this case we use the Krylov-Bogoliubov (KB) method \cite{bogol} for our analysis.
The KB {\em Ansatz} is based on the assumption that the CNT displacement
takes the form $u(t)$$=$$A(t)$$\sin(\omega_0 t$$+$$\psi(t))$, where both the
amplitude $A(t)$ and the phase $\psi(t)$ are slowly varying
functions of time and the CNT mechanical energy is $m\omega_0^2A^2/2$. Substituting into Eq.~(\ref{newteq1}), multiplying by $\dot u(t)$, and
averaging over one oscillation period %period of the mechanical oscillations
we get an equation for the rate of change of $A(t)$,
%%%%%%%%%%%%%%%%%%%%%%%%%%%%%%%%%%%%%%%%%%%%%%%%%%%%%%%%%%%%%%%%%%%%%%%%%%%%%%%%%%%%%%%%
\begin{eqnarray}\label{amp}
&&\frac{d A^2}{dt}=\omega_0\left[W_1(A)-W_0(A)\right],
\\
&&W_1(A)=-\frac{A^2}{\pi}\left(u_0\frac{2F\nu_h}{
\Gamma^2}\right)^2\int_0^{2\pi}\cos^2(\psi)f_2[A\sin(\psi)]d\psi,\nonumber
\end{eqnarray}
where $W_1(A)$ is proportional to the work done on the NEM system by the magnetic force 
during one oscillation period and $W_0(A) = A^2/Q_0$ is proportional to the energy dissipated due 
to intrinsic friction during the same time; the solution $W_1(A)$$=$$W_0(A)$ yielding
a stationary solution.
In Fig.~2 the ratio $W_1(A)$$/$$W_0(A)$ is plotted as a function of $A$ 
for three different microwave intensities ($\nu_h$), corresponding to three
%From this figure one can distinguish three
different regimes. 
Regime (i) occurs at low microwave intensities  when the work
done by the magnetic force is smaller than the dissipated energy ($W_1/W_0$$<$$1$) for any oscillation amplitudes and
hence the only stationary solution is static, $A$$=$$0$.
Regime (ii) arises at intermediate radiation intensities where
%the ``intrinsic-dissipation" function
%$W_1(E)=W_{0}(E)$ %intersects the function $W_1$$=W_1(E)$
$dA^2/dt$$=$$0$ for three different amplitudes, each corresponding to a stationary solution.
Two of these solutions, one static and one with finite oscillation amplitude, are
stable while the third (finite-amplitude) solution is unstable. Note that if the NEM system is initially
static, it will remain so in this regime. The third regime (iii) emerges at high intensities of the microwave radiation
where the intrinsic mechanical dissipation overcomes  the work
performed by magnetic forces only at a certain finite mechanical vibration amplitude $A_{\rm lim}$, 
which defines the amplitude of the CNT self-sustained oscillations.
Together, the intensity regimes (i)-(iii) determine a so called hard instability scenario 
%{\color{blue}All mentioned above regimes determine a scenario of a hard instability} 
\cite{Isaacs}, which results in a hysteretic behaviour of the system characteristics as a function of the microwave field intensity $\nu_h$. 

The self-sustained oscillations, being due to electronic spin-flip transitions, will generate a correction $\delta I $$\propto$$ A_{\rm lim}^2$ to the electrical current through the NEM system, %$\delta I $$\propto$$ A_{\rm lim}^2$, 
which can be detected. An estimation using the system parameters given in Fig.~2 shows that $\vert \delta I \vert $$\sim$$ 10-100\,$pA. 
Alternatively the CNT bending vibrations can be detected by using standard methods such as the rectification technique of Ref.~\onlinecite{zantnano} or the scanning force microscopy technique of Ref.~\onlinecite{sfm}.

\begin{figure}
\centering
\includegraphics[width=0.8\columnwidth]{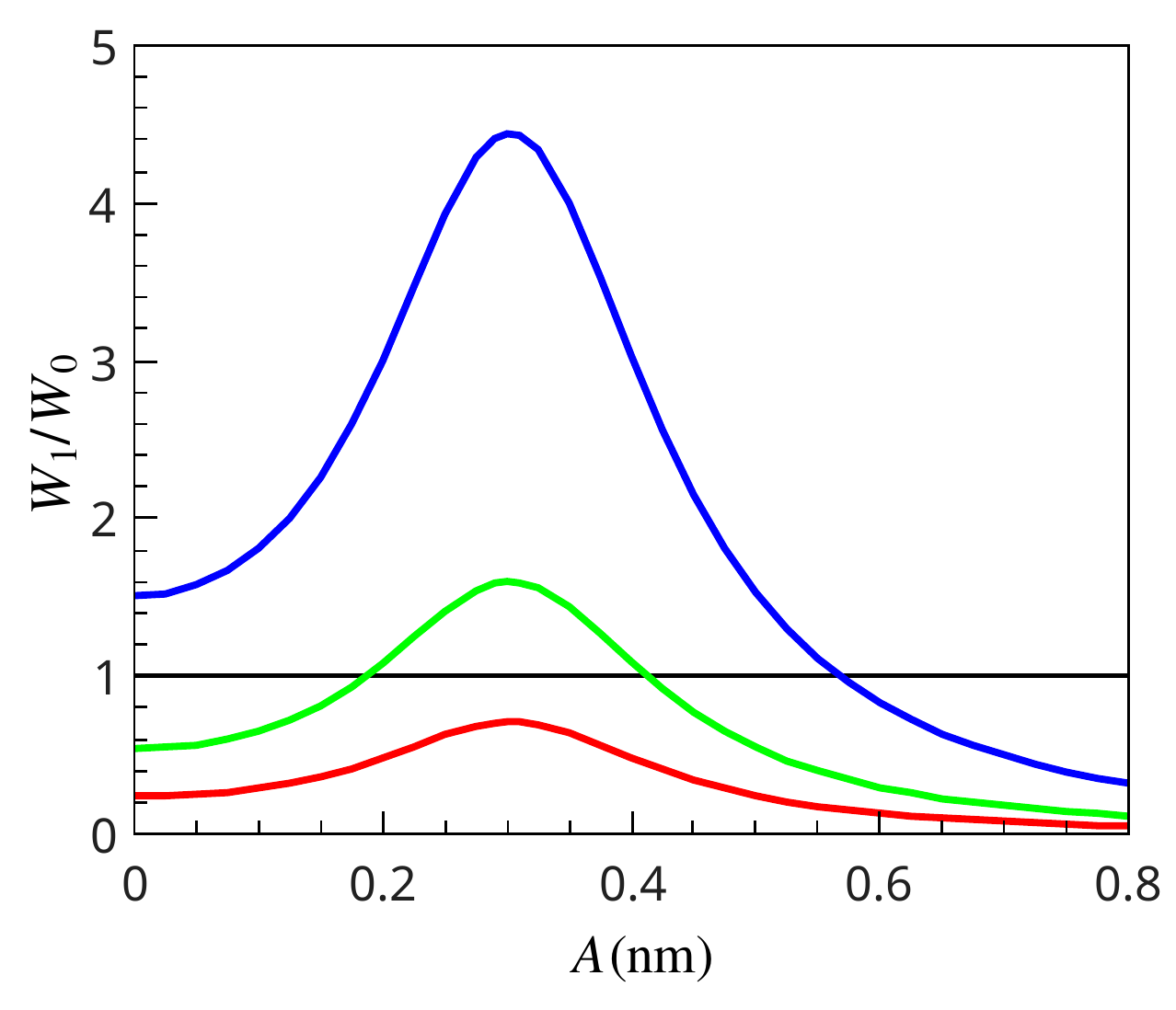}
\caption{
Ratio between work done on the NEM system by the magnetic force ($\propto$$W_1$) and energy dissipated due to intrinsic friction ($\propto$$W_0$) plotted as a
function of CNT oscillation amplitude $A$ for 
%{\color{blue}$\Gamma =0.6\,$GHz} 
$\Gamma =0.6\cdot 10^9$ s$^{-1}$, %[rad/s ?]}, 
$\Delta$$-$$\Omega$$ =$$ \Gamma$, $u_0$$=$$2\,$pm, $\partial $$B_z/$$\partial z$$ =$$5 $$\cdot$$10^6\,$T/m, $Q_0$$ =$$10^6$, and three different microwave field intensities $\nu_h$.  
%{\color{blue}The value $W_1/W_0=1$ defines a non-trivial ($A$$\neq$$0$) stationary solution of Eq.~(\ref{amp}).}
If $W_1/W_0<1$ for some $A$, then $A$ decreases with time according to Eq.~(\ref{amp}), while if $W_1/W_0>1$, $A$ increases with time. 
If $W_1/W_0=1$ (black horizontal help-line) for some finite amplitude, then Eq.~(\ref{amp}) has a nontrivial ($A \neq 0$) stationary solution for that amplitude. 
The red (bottom) curve ($\nu_h$$=$$0.4\,\Gamma$) describes the
behaviour of the NEM system below the instability threshold and
illustrates the text's regime (i), which has no non-trivial stationary solution. %{\color{blue}described in the text}. 
The green (middle) curve ($\nu_h$$=$$0.6\,\Gamma$) corresponds to regime (ii), where the mechanical system is characterized by %{\color{blue}three} 
two non-trivial stationary solutions. 
The blue (top) curve ($\nu_h$$=$$1.0\,\Gamma$) corresponds to regime (iii), 
%{\color{blue}where a mechanical instability leads to self-sustained oscillations of the resonator.} 
with a single non-trivial stationary solution to Eq.~(\ref{amp}) for $A=A_{\rm lim}$. In this regime a mechanical
instability leads to self-sustained resonator oscillations of amplitude $A_{\rm lim}$ \cite{Robcom}. 
}
\label{Fig}
\end{figure}

In conclusion we have suggested a new spin-mediated
photo-mechanical mechanism for coupling electronic and mechanical degrees of freedom
in a nano-electro-mechanical (NEM) device comprising a carbon nanotube (CNT) resonator suspended between to magnetic electrodes and under the influence of an inhomogeneous magnetic field from a nearby %STM
tip-shaped magnetic gate.
The predicted strong effect of microwave radiation on the nanomechanics of the device
is based on two features: (i) the resonant nature of the microwave-induced electronic spin-flip transitions in the CNT resonator, and (ii) a latent nano-mechanical instability of the resonator caused by a spin-dependent magnetic force that pumps energy into its vibrations.
The instability occurs if the pumping rate [see Eq.~(\ref{qual})] exceeds the dissipation rate, set by the quality factor $Q_0$ of the resonator.
We find that this criterion leads to a lower bound of about 1~mT for the amplitude of the magnetic component ($h$) of the microwave field.

The predicted photo-induced mechanical instability develops into pronounced vibrations of the CNT resonator. These are accompanied by temporal oscillations of the spin accumulated in the CNT and related significant effects on the spin-dependent electrical current through the device corresponding to a highly efficient (up to 100\%) photo-electric transduction effect. The results obtained for the studied CNT-based magnetic NEM device provide a basis for pursuing further spintronics applications, including but not limited to spin-current rectifiers (filters, splitters, ratchets, etc.), and as elements of ac- and dc- circuits and devices with controllable (non-linear) damping. 

%\vspace*{-10mm}
We acknowledge fruitful discussions with A. H{\" u}ttel and S. Ludwig.
This work was partially supported by the Swedish Research
Council (VR).

%\vspace*{-5mm}


\begin{thebibliography}{99}
\bibitem{footnote2}
For asymmetric tunnel couplings and optimal resonance conditions, $\vert\Omega$ $-$  $\Delta\vert$
$\sim$ $\Gamma_D$, 
$\Gamma$ in Eq. (\ref{qual}) should be replaced by $\Gamma_D$ ($\Gamma_D$ $\ll$ $\Gamma_S$)
or $\sqrt{\Gamma_S\Gamma_D}$ ( $\Gamma_D$ $\gg$ $\Gamma_S$). This will not qualitatively change the NEM
instability criterion.
\bibitem{directions} Note that the direction of the magnetic moment of
an electron is opposite to that of its spin.
\bibitem{Robcom} Typical values of the amplitudes of the developed mechanical vibrations are set by
the distance between the point defined by the static solution: $A=0$ and emergent point where the spin flip assisted resonant absorption of the microwave field is achieved.
\end{thebibliography}
\end{document}